\documentclass[reprint,
 superscriptaddress,
 amsmath,amssymb,
 aps,
 prb
]{revtex4-2}
	\usepackage[T1]{fontenc}
	\usepackage[utf8]{inputenc}
    \usepackage[pdftex,bookmarks=false]{hyperref}
    \usepackage{microtype}
    \usepackage{graphicx}
    \usepackage{dcolumn}
    \usepackage{bm}
    \usepackage{multirow}
    \usepackage{booktabs}
    \usepackage{xcolor}
    \usepackage{mathtools}
    \usepackage{mathptmx}
    \usepackage{upgreek}
    \usepackage{float}
    \usepackage{siunitx}

    \hypersetup{
    	colorlinks = true, 
		citecolor = blue,
		bookmarksnumbered = true,
		urlcolor = green
	}
	
    \DeclarePairedDelimiterX\braket[2]{\langle}{\rangle}{#1 \delimsize\vert #2}

    \renewcommand{\Re}{\operatorname{Re}}

\definecolor{green}{rgb}{0,0.6,0.1}

\usepackage[normalem]{ulem}

\usepackage[caption=false,position=top,captionskip=0pt,farskip=0pt]{subfig}
\begin{document}

\title{Magneto-optical signatures of plateau states in the kagome spin-ice HoAgGe}
\author{F.~Schilberth}
\email{felix.schilberth@physik.uni-augsburg.de}
\affiliation{Experimentalphysik V, Center for Electronic Correlations and Magnetism, Institute for Physics, Augsburg University, D-86135 Augsburg, Germany}
\affiliation{Department of Physics, Institute of Physics, Budapest University of Technology and Economics, M\H{u}egyetem rkp. 3., H-1111 Budapest, Hungary}

\author{L.~DeFreitas}
\affiliation{Department of Physics,
Colorado State University, Fort Collins, CO, USA}
\affiliation{School of Materials Science and Engineering, Colorado State University, Fort Collins, CO, USA}

\author{J.~Guo}
\affiliation{School of Physics, Beihang University, Beijing 100191, China}

\author{N.~Herrmann}
\affiliation{Experimentalphysik V, Center for Electronic Correlations and Magnetism, Institute for Physics, Augsburg University, D-86135 Augsburg, Germany}

\author{F. LeMardelé}
\affiliation{Laboratoire National des Champs Magnétiques Intenses, LNCMI-EMFL, CNRS UPR3228, Univ. Grenoble Alpes, Univ. Toulouse, Univ. Toulouse 3, INSA-T, Grenoble and Toulouse, France}

\author{I. Mohelsky}
\affiliation{Laboratoire National des Champs Magnétiques Intenses, LNCMI-EMFL, CNRS UPR3228, Univ. Grenoble Alpes, Univ. Toulouse, Univ. Toulouse 3, INSA-T, Grenoble and Toulouse, France}

\author{M. Orlita}
\affiliation{Laboratoire National des Champs Magnétiques Intenses, LNCMI-EMFL, CNRS UPR3228, Univ. Grenoble Alpes, Univ. Toulouse, Univ. Toulouse 3, INSA-T, Grenoble and Toulouse, France}

\affiliation{Faculty of Mathematics and Physics, Charles University, Ke Karlovu 5, Prague, 121 16, Czech Republic}

\author{U.~Nagel}
\affiliation{National Institute of Chemical Physics and Biophysics, 12618 Tallinn, Estonia}

\author{T.~R$\tilde{\text{o}}\tilde{\text{o}}$m}
\affiliation{National Institute of Chemical Physics and Biophysics, 12618 Tallinn, Estonia} 

\author{P.~Gegenwart}
\affiliation{Experimentalphysik VI, Center for Electronic Correlations and Magnetism, Institute for Physics, Augsburg University, D-86135 Augsburg, Germany} 

\author{H.~Chen}
\affiliation{Department of Physics,
Colorado State University, Fort Collins, CO, USA}
\affiliation{School of Materials Science and Engineering, Colorado State University, Fort Collins, CO, USA}

\author{K.~Zhao}
\affiliation{School of Physics, Beihang University, Beijing 100191, China}
\affiliation{Experimentalphysik VI, Center for Electronic Correlations and Magnetism, Institute for Physics, Augsburg University, D-86135 Augsburg, Germany} 

\author{I.~K\'ezsm\'arki}
\affiliation{Experimentalphysik V, Center for Electronic Correlations and Magnetism, Institute for Physics, Augsburg University, D-86135 Augsburg, Germany}

\author{S.~Bord\'acs}
\affiliation{Experimentalphysik V, Center for Electronic Correlations and Magnetism, Institute for Physics, Augsburg University, D-86135 Augsburg, Germany}
\affiliation{Department of Physics, Institute of Physics, Budapest University of Technology and Economics, M\H{u}egyetem rkp. 3., H-1111 Budapest, Hungary}
\affiliation{HUN-REN-BME Condensed Matter Research Group, Budapest University of Technology and Economics, M\H{u}egyetem rkp. 3., H-1111 Budapest, Hungary}

\begin{abstract}
Coupling of itinerant electrons to non-collinear spin structures in frustrated rare-earth magnets leads to unusual transport properties. While these complex magnetic orders are expected to modify the electronic structure, underlying microscopic mechanisms have been rarely identified. Using magnetoreflectance spectroscopy and \textit{ab initio} calculations, we reveal reconstructions of the electronic structure across the magnetization plateaus of the kagome spin-ice HoAgGe. We find an unusual optical anisotropy: a larger conductivity perpendicular to the kagome planes originating from the hybridization of Ho $4f$ flat bands with conduction electrons. Moreover, these flat bands are also responsible for the pronounced magnetoreflectance observed at the metamagnetic transitions. Most importantly, the magnetoreflectance behaves completely differently for excitations near the Fermi level and higher-energy interband transitions. The former, expected to sensitively probe variations in band topology near the Fermi level, shows abrupt non-monotonous changes between the magnetic states, while the interband excitations mimic the field dependence of the magnetization. These observations imply that in HoAgGe, the unusual magnetotransport likely reflects variation of band topology between plateau states. More generally, our study shows that heavy-electron bands near the Fermi level offer an efficient way to control the band structure and thereby the transport and optical properties of non-collinear magnets.
\end{abstract}
\maketitle

\section{Introduction}

Permanent magnets containing rare-earth elements are the most frequently used magnets in daily technologies, due to their large local moments, associated with electrons on the $4f$ shell, and strong anisotropy \cite{Yang2016}. When interactions compete between these moments, peculiar magnetic states emerge, including spin-ice states on geometrically frustrated lattices \cite{Bramwell2001, Yasui2001, Singh_2008, Morris2009, Zhao2020, Ortiz2024}, non-collinear and non-coplanar magnetic structures \cite{Kurumaji2019, Hirschberger2019, Khanh2020, Zhao2020, Simeth_2024}. These exotic localized spin textures can generate a Berry curvature acting on the charge carriers and also modify their mass and/or scattering rate \cite{Taguchi2001, Kurumaji2019, Hirschberger2019, Khanh2020, Zhao2024, Chen2025}. 

Rare-earth compounds with pyrochlore and kagome lattices provide a plethora of magnetic phases for investigating this interplay between localized spin textures and electronic degrees of freedom.
Pyrochlore magnets can realize, for instance, the magnetic analogue of water ice, a disordered frozen state formed with 2--in--2--out spin configuration on each tetrahedron \cite{Bramwell2001}. Systems built of kagome layers also host a large variety of magnetic orders, as exemplified in the RMn$_6$Sn$_6$ or Fe$_x$Sn$_y$ series \cite{Yin2020, Kang2019}. They can potentially realize another spin-ice state, with 2-in-1-out and 1-in-2-out spin configurations on the triangular plackets forming the kagome layers. In both pyrochlore and kagome compounds, the exchange coupling of the largely localized $4f$ electrons and their hybridization with itinerant electrons can produce pronounced anomalous and topological Hall effects as well as non-monotonic magnetoresistance \cite{Taguchi2001, Hirschberger2019, Zhao2024}. However, due to the various possible contributions, including intrinsic Berry-curvature related terms and extrinsic ones originating from spin-dependent scattering, it is notoriously difficult to identify the underlying microscopic mechanisms. The central question is therefore how complex spin orders of $4f$ moments affect the electronic structure, and, in turn, the charge transport and optical properties. 

HoAgGe offers an ideal laboratory to address this question, as it exhibits a cascade of magnetic field-driven phase transitions between states all obeying the kagome spin-ice rules, owing to the Ising character of the Ho $4f$ spins. Here, we choose magneto-optical spectroscopy to tackle with this problem due to its capability  to probe electronic states in a spin-sensitive and energy-resolved fashion. In HoAgGe, the rare-earth Ho atoms form a distorted kagome lattice with up-- and downward facing triangles rotated by 15.6$^\circ$ in opposite directions in the $ab$ plane, resulting in the hexagonal space group $P\bar{6}2m$ \cite{Morosan2004,Zhao2020} \footnote{The following notation is used in our work: The $a$ axis is defined as one of the hexagonal axes, with $b$ denoting the perpendicular direction in the kagome plane and $c$ is the stacking direction.}. Recent neutron diffraction results revealed that HoAgGe exhibits a coplanar antiferromagnetic ordering of the Ho spins below $T_1=\SI{11.6}{K}$ \cite{Zhao2020, Zhao2026} which breaks lattice translational symmetry while retaining fluctuating magnetic charges, as evidenced by persistent diffuse scattering \cite{Zhao2026}. Upon cooling below $T_2=\SI{7}{K}$, these fluctuations are progressively suppressed, leading to an ordered ground state with $\pm1$ magnetic monopole charges on every triangular unit due to the 2 in-1 out or 1 in-2 out kagome ice constraint \cite{Wills2002}.

HoAgGe undergoes a series of metamagnetic transitions for external magnetic fields applied along the $a$ or $b$ axis of the crystal, while it can show a spin--flop transition for field along $c$, similar to other members of the RAgGe (R = rare earth) family \cite{Morosan2004,Zhao2020, Zhao2024}. These metamagnetic states appear as plateaus in the magnetization curve for magnetic field along the easy magnetization axis, which is the $b$ axis in HoAgGe \cite{Zhao2020}. The evolution of the magnetic order consists of the ground state, the 1/3 and 2/3 magnetization plateaus and the saturated state, respectively. Due to the strong Ising character of the rare-earth moments, for each transition, 1/3 of the spins flip into the direction of the field. Though this process increases the magnetization, every plateau state still obeys the kagome ice constraint \cite{Zhao2020, Zhao2024}. Importantly, the metamagnetic transitions between the plateau states in HoAgGe are accompanied by sharp anomalies in magnetotransport. 
The longitudinal resistivities, $\rho_{aa}$ and $\rho_{cc}$, as well as the Hall resistivity $\rho_{ac}$ show strongly non--monotonous evolution between the magnetization plateaus \cite{Zhao2024, Li2022, Zhao2024}. Based on the magnetotransport anomalies, substantial magnetic-order-induced changes in the band structure were inferred, including variations of band topology \cite{Zhao2024, Zhao2026, Bhandari2025}.

In order to gain insight into the nature of the electronic states near the Fermi level, giving rise to the highly unusual magnetotransport properties, we investigated the band structure of HoAgGe by polarized optical and magneto-optical spectroscopy, combined with \textit{ab initio} calculations.
We observe a strong anisotropy of the optical properties: The optical conductivity is larger for light polarization perpendicular to the kagome plane than parallel to it, disproving the applicability of the single-kagome-layers description for this material. Our central result is the observation of drastic changes in the reflectivity upon the metamagnetic transitions, and the striking difference in its field dependence: non-monotonous changes in the range of the free-carrier response, contrasted by a monotonous variation at interband transitions. The former follows the field dependence of the dc magnetoresistance, while the latter mimics  the field dependence of the magnetization. Our material specific theory demonstrates that unoccupied Ho $4f$ flat bands hybridized with $p$ and $d$ orbitals of Ag and Ge govern the interband transitions in HoAgGe, leading to the optical anisotropy and the strong magnetoreflectance, while the peculiar magnetotransport is likely associated to changes in the electronic band topology between the plateau phases.

\begin{figure}
    \centering
    \includegraphics[width=\linewidth]{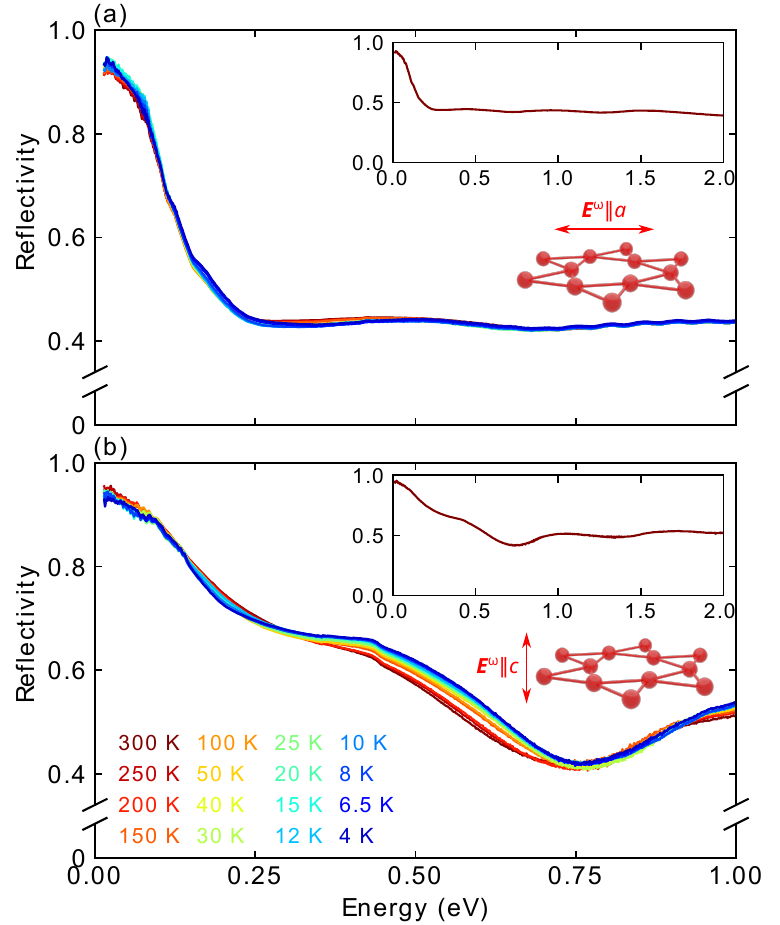}
    \caption{Polarized reflectivity spectra of HoAgGe. (a) In--plane and (b) out--of--plane reflectivity spectra at various temperatures. The insets display the light polarization in the two cases as well as the room-temperature reflectivity spectra over the full experimental range. }
    \label{fig:Reflectivities}
\end{figure}

\section{Results}
\subsection{Zero-field reflectivity}
We carried out polarized magneto-reflectance experiments on polished $ab$, $ac$ and $bc$-cut crystals with diameters on the mm scale. The in--plane reflectivity spectra (${E}^\omega\parallel a$) and the out--of--plane reflectivity spectra (${E}^\omega\parallel c$) are shown in Fig.\,\ref{fig:Reflectivities}(a) and (b), respectively. For both polarizations, the metallic character is evident as the reflectivity approaches one towards low frequencies. For the in--plane spectra, we obtain the plasma edge around \SI{120}{meV}, followed by an almost constant reflectivity of approx. 40\% above \SI{250}{meV}. In this polarization, the reflectivity has only a weak temperature dependence. The out--of--plane reflectivity spectra above the plasma edge exhibit more features: A shoulder that is followed by a minimum at \SI{750}{meV}, both getting more pronounced at low temperatures. At higher energies, an additional local minimum develops around \SI{1.4}{eV} above which no temperature dependence was observed. These spectra therefore reveal a strong optical anisotropy.

\begin{figure}[t]
    \centering
    \includegraphics[width=\linewidth]{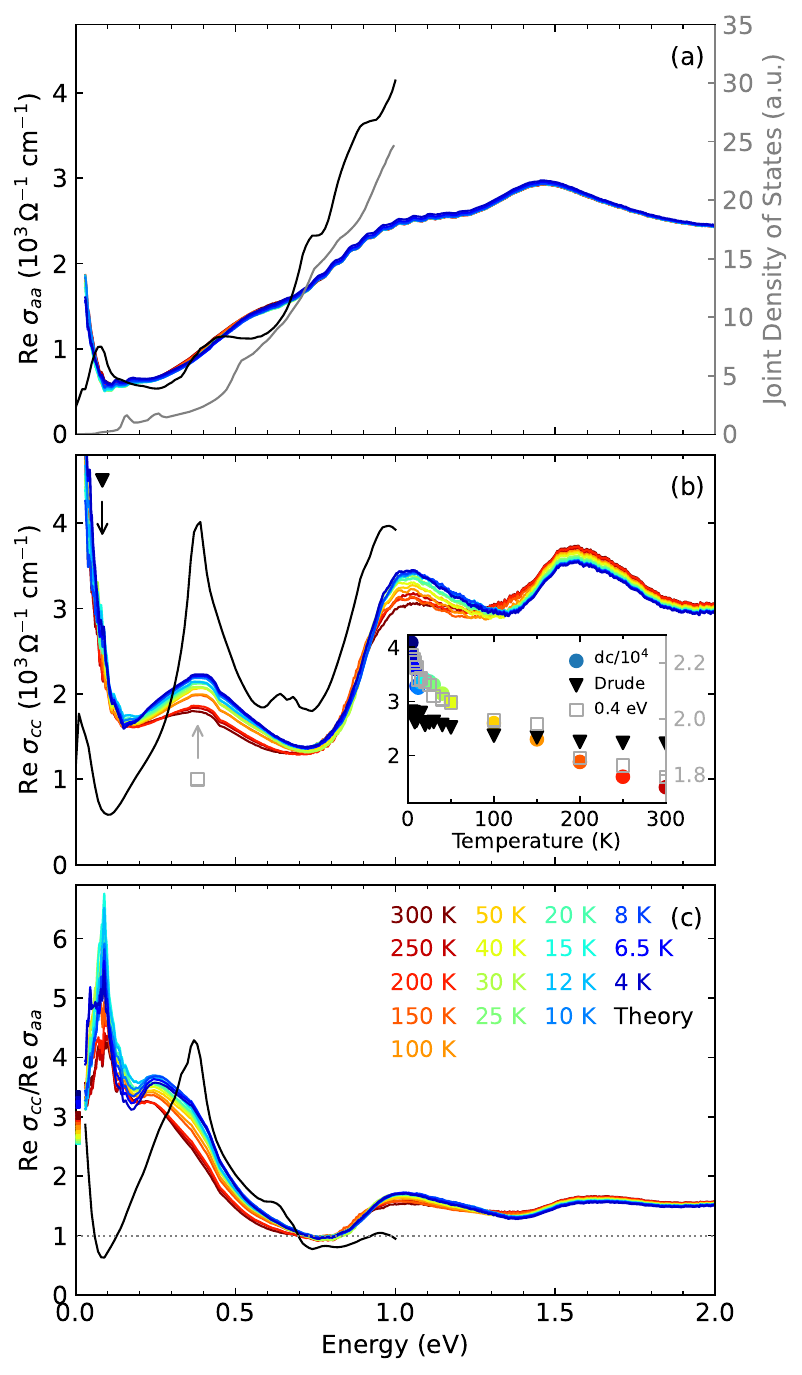}
    \caption{Optical conductivity data for HoAgGe. Panels (a) and (b) show the in--plane and out--of--plane optical conductivity, $\Re\,\sigma_{aa}$ \& $\Re\,\sigma_{cc}$ respectively, while panel (c) displays the optical anisotropy $\Re\,\sigma_{cc}/\Re\,\sigma_{aa}$. The dc values are shown as squares at zero energy for the anisotropy only as the dc conductivity values lie above the range shown here. For comparison, the spectra obtained from the \textit{ab initio} calculations are shown in black for all quantities up to \SI{1}{eV}, as well as the joint density of states in grey in panel (a).}
    \label{fig:Conductivities}
\end{figure}

\subsection{Zero-field optical conductivity}
For a more quantitative analysis and comparison with the optical response calculated \textit{ab initio}, we derived the real part of the optical conductivity spectra, as shown for both polarizations at various temperatures in Fig.\,\ref{fig:Conductivities}. At energies above \SI{0.2}{eV}, the in--plane conductivity $\Re\,\sigma_{aa}$ in panel (a) exhibits two shoulders at approx. \SI{0.4}{eV} and \SI{1}{eV} and increases monotonously up to a broad maximum around \SI{1.5}{eV}. The spectra are almost independent of temperature. By contrast, the out--of--plane conductivity $\Re\,\sigma_{cc}$ in Fig. \ref{fig:Conductivities}(b) shows strong temperature dependence and the features observed in the other polarization emerge as well-pronounced peaks centered around 0.4, 1 and \SI{1.5}{eV}, respectively. These are interband transitions whose nature will be discussed later. The temperature dependence of the Drude term together with that of the 0.4 eV peak is displayed in the inset in comparison with the dc conductivity. For both $\sigma_{aa}$ and $\sigma_{cc}$, we observe the foot of a Drude component in the $\omega\rightarrow0$ limit. Note that the Drude term, which describes the free carrier dynamics, is rather steep and the dc conductivity values are high reaching 4100 and \SI{12600}{\per\ohm\per\cm} already at \SI{300}{K} for the in-- and out--of--plane directions, respectively.

Importantly, both the dc conductivity and the optical conductivity over the spectral range studied here is larger perpendicular to the kagome layers than parallel to them. This implies a strong coupling between the individual kagome planes, which is consistent with the short $c$-axis lattice constant, and a three-dimensional electronic structure \cite{Zhao2020,Zhao2024, Morosan2004}. In fact, an interlayer conductivity larger than the intralayer one is not uncommon for kagome metals as it was reported for, e.g., YCr\textsubscript{6}Ge\textsubscript{6} \cite{Yang2022} and the Fe$_x$Sn$_y$ family \cite{Du2022, Sales2019, EbadAllah2024}. For HoAgGe, the dominance of the interlayer conductivity persists also at finite frequencies, with the ratio $\Re\,\sigma_{cc}/\Re\,\sigma_{aa}$ exceeding one for most of the covered energy range, and converges to the static value towards zero frequency, as shown in Fig.\,\ref{fig:Conductivities}(c).

\subsection{\textit{ab initio} analysis}
To reveal the character of the pronounced interband transitions, we have studied the electronic structure and optical conductivities of HoAgGe by \textit{ab initio} calculations in the antiferomagnetic ground state. 

\begin{figure}[h]
	\centering
	\includegraphics[width=\linewidth]{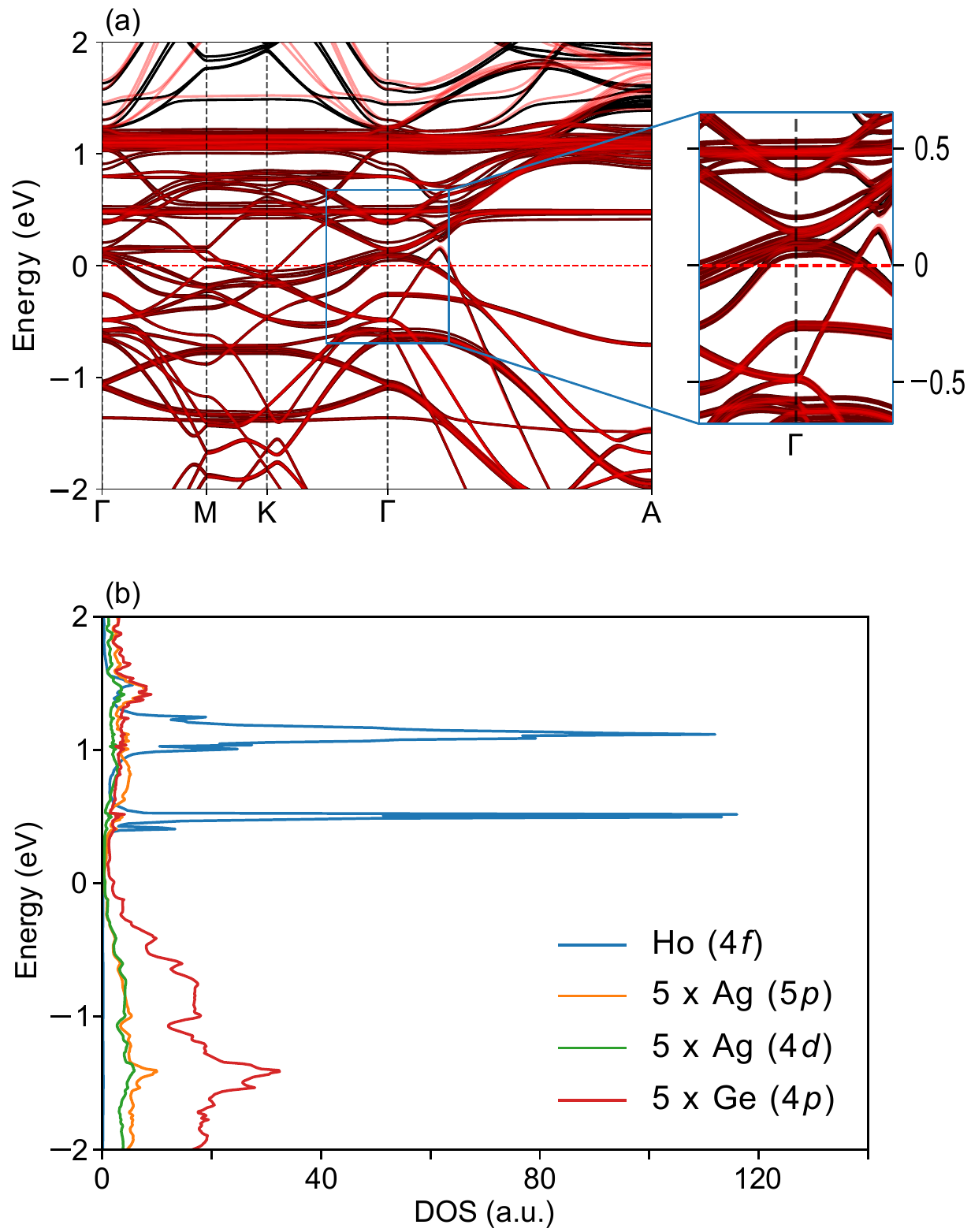}
	\caption{DFT results of HoAgGe in the ground state. (a) DFT (black solid lines) and Wannier-interpolated (red solid lines) band structures near the Fermi energy $E_\text{F}$ chosen as 0. (b) Orbital-projected density of states. The Ag and Ge species are multiplied by a factor of 5 to enhance their visibility. } 
\label{fig:DOS}
\end{figure}

The obtained Wannier-interpolated band structure (red solid lines) agrees reasonably well with the DFT result (black solid lines) up to the upper bound of the frozen window, i.e. up to $\sim \SI{1}{eV}$ above the Fermi energy, in Fig.~\ref{fig:DOS}(a). Fig.~\ref{fig:DOS}(b) displays the orbital-resolved density of state (DOS) from the DFT calculations over the same energy range. The two prominent peaks above the Fermi level originate from the $4f$ orbitals of Ho and correspond to the flat bands in panel (a). It should be noted that in itinerant systems like HoAgGe such unoccupied Kohn-Sham orbitals do not correspond to low-lying excited states of a Ho$^{3+}$ ion defined by spin-orbit and crystal electric field couplings, as observed experimentally in insulating compounds \cite{Dieke1963}. Nevertheless, Ho $4f$ states strongly affect optical properties in the way elaborated below.

The joint DOS (JDOS) and the real part of the optical conductivities in both polarizations are plotted in Fig.\,\ref{fig:Conductivities} in comparison with the experimental optical conductivity spectra up to \SI{1}{eV}, i.e. over the energy range, where the Wannier-interpolated and DFT band structures are consistent. First of all, the calculated conductivity spectra reproduce the peak- and shoulder-like features, respectively observed in the out--of--plane and in--plane conductivity spectra above the region of the Drude peak. The Drude response of free carriers was not obtained by Wannier90. Moreover, the calculated in--plane optical conductivity $\sigma_{aa}$ closely follows the JDOS, suggesting that the electric dipole matrix element in this polarization has no significant energy dependence. More precisely, the matrix element is not specific to the initial and final states involved in these transitions. In contrast, $\sigma_{cc}$ strongly deviates from the JDOS trend, in particular, the well-separated interband excitations appear at 0.4 and \SI{1}{eV}. We therefore conclude that the strong optical anisotropy is the consequence of the large, energy-dependent out-of-plane component of the electric dipole matrix element, $r_c$, while the in-plane component $r_a$ is smaller but also energy independent, at least for occupied and empty states separated by not more than \SI{2}{eV}.

In the following, we discuss the possible origin of the interband transitions.
The band structure and the DOS projected to the different atomic orbitals Fig.~\ref{fig:DOS} (a) and (b), respectively, suggest the presence of two sets of $4f$ flat bands located at $\sim 0.5$ and \SI{1.1}{eV}. Thus, not even the former one can directly contribute to the $\sigma_{cc}$ peak at $\sim \SI{0.4}{eV}$, but indirectly via hybridization with dispersive Ag and Ge orbitals. As seen in the inset of Fig.\,\ref{fig:DOS}(a), some of the dispersive bands near $\Gamma$ are split by  $\sim\SI{0.4}{eV}$ and have very similar dispersions, a condition leading to large contribution to the JDOS. The pronounced $r_c$ matrix element between these bands suggest that they are split by a local ($\bm k$-independent) potential term that does not commute with $r_c$ in the subspace of these bands, and therefore mostly depends on $r_a$ and $r_b$. Moreover, we note that the CEF Hamiltonian of Ho$^{3+}$ in the $^5I_8$ ground state has a similar energy scale and a dominant in-plane quadrupole term \cite{Zhao2020}. Taken together, the above evidence suggests that the \SI{0.4}{eV} peak of $\sigma_{cc}$ is due to dispersive bands of $s,p,d$ character near $E_\text{F}$ split by their Coulomb coupling with the $4f$ electrons of Ho$^{3+}$ in a mean-field manner.

A more direct contribution of $4f$ flat bands to the \SI{1}{eV} peak of $\sigma_{cc}$, as final states, is implied by their dominance in the DOS above the Fermi energy, as see Fig.\,\ref{fig:DOS}(b).
These higher $4f$ excited states are weakly hybridized with Ag and Ge orbitals at similar energies.
Due to the $4f$ states, one expects a pronounced matrix element effect, similarly to the case of the \SI{0.4}{eV} peak, that becomes progressively suppressed as temperature increases towards the CEF scale. This is in agreement with the strong temperature dependence of the out--of--plane optical spectra observed below approx.~\SI{50}{K} in the range of the three interband transitions in the inset of Fig.\,\ref{fig:Conductivities}(b). The low-temperature increase of the optical anisotropy indicates further distortion of the charge distribution, which via the strong spin-orbit coupling of the rare-earth modifies the magnetic anisotropy crucial in the description of this spin-ice compound. As we will show later, this can be further corroborated by the strong magneto-optical response observed in this spectral range, which can be similarly enhanced by a modification of the transition matrix elements to the $4f$ levels due to the magnetic state.

\subsection{Magnetoreflectance}
\begin{figure*}
	\centering
	\includegraphics[width=\linewidth]{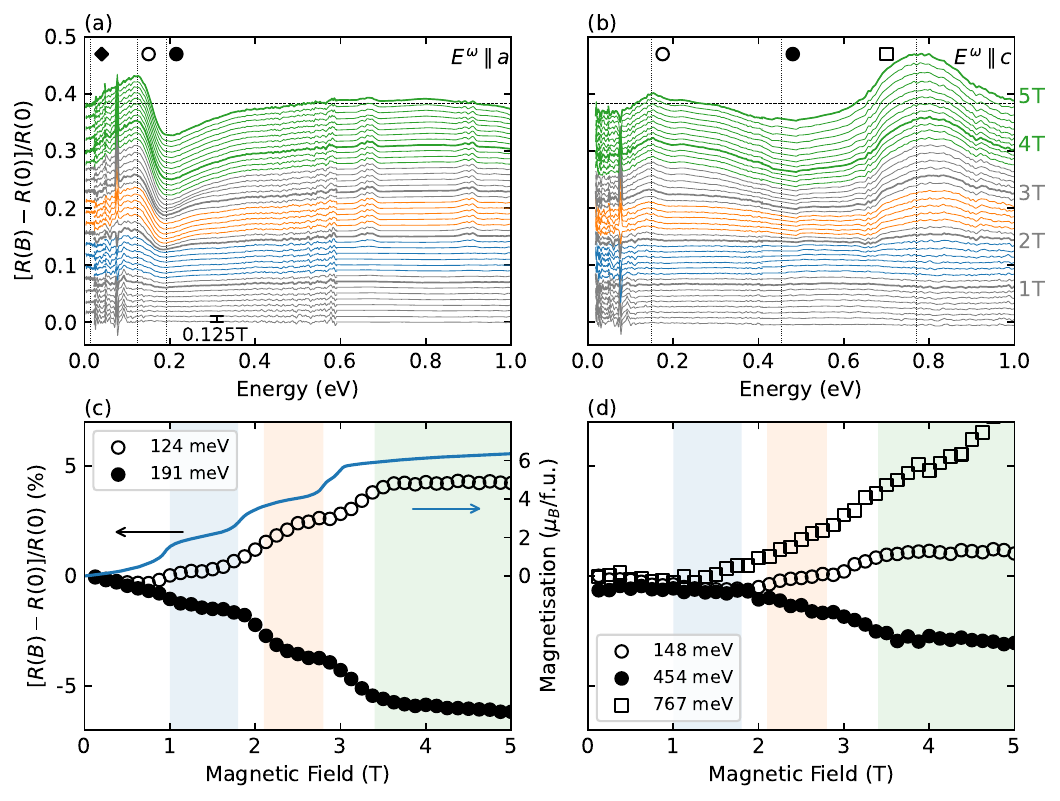}
	\caption{Magnetoreflectance data for HoAgGe with $H\parallel b$. (a, b) Magnetoreflectance spectra for the light electric field polarized along the $a$ and $c$ crystallographic axis, respectively. Spectra were taken from $0-\SI{5}{T}$ with \SI{0.125}{T} steps. Subsequent spectra are shifted vertically in proportion to the field value and are colored according to the background shading for field ranges of the plateaus in the magnetization in panel (c) (reproduced from Ref.~\cite{Zhao2024}). The horizontal dashed line represents the zero for the \SI{5}{T} spectrum to highlight the relative changes. (c, d) Field dependence extracted at fixed photon energies, indicated by the vertical dashed lines and associated symbols in panels (a,b). For in--plane light polarization, the reflectance shows distinct plateaus at the metamagnetic phases, following the behavior of the magnetization at \SI{5}{K}.
	}
	\label{fig:Magnetoreflectance}
\end{figure*}
As the main target of our study, in the following, we investigate the effect of metamagnetic transitions on the electronic structure of HoAgGe via the corresponding variations of the optical response, in the spectral range of the lowest two interband transition located at around 0.4 and \SI{1}{eV} as well as the onset of the Drude response. For this, we measured the change of the reflectance induced by the applied magnetic field (i.e. the magnetoreflectance) for $H\parallel b$ and with the electric field of light polarized along the crystallographic $a$ and $c$ axes is shown in Fig. \ref{fig:Magnetoreflectance} (a) and (b), respectively. For the in--plane polarization in panel (a), the magnetoreflectance approaches zero at the high- and low-frequency cutoffs. In between, a dispersive spectral feature dominates the magnetoreflectance: Around \SI{124}{meV} a peak develops with increasing magnetic field, followed by a minimum around \SI{200}{meV}. The detailed field dependence of the minimum and maximum is shown in Fig. \ref{fig:Magnetoreflectance}(c) with \SI{125}{mT} field resolution. Upon approaching the 1/3 and 2/3 magnetization plateaus, both the maximum and the minimum of the  magnetoreflectance spectrum exhibit sudden changes, and levels off at a value as high as $\pm5\%$ in the saturated state. The monotonous step-like evolution of the magnetoreflectance for ${E}^\omega\parallel a$ mimics the field dependence of the magnetization plotted on the second $y$ axis. The field ranges of the different plateau phases are indicated by the background shade. Since the magnetoreflectance measurement was carried out at 4\,K, the magnetic transitions are slightly smeared out and the 1/6 and 5/6 plateau states do not show any distinct changes. This is in line with the magnetization data at this temperature \cite{Zhao2024}. 

The close correspondence between the field dependence of the magnetoreflectance and the magnetization and the strong contrast to the non-monotonous behaviour of the dc magnetotransport data implies that the magnetoreflectance is dominated by the lowest interband transitions and not by the response of the free carriers to the magnetic field, even at energies down to \SI{100}{meV}. This is corroborated by the optical conductivity, from which we can estimate a width of the Drude peak below \SI{10}{meV} at low temperatures, far below the frequencies of the observed features in the magnetoreflectance.

Before we turn to the consequences of this observation, we also discuss the magnetoreflectance spectra for out--of--plane polarization ${E}^\omega\parallel c$, shown in Fig.\,\ref{fig:Magnetoreflectance}(b). Again, a peak and dip emerge for increasing field, but the spectral shape is different and the position of these features are shifted to higher energies. The peak now appears at \SI{150}{meV} which is the bottom of the Drude peak in the optical conductivity, while the dip is shifted to \SI{450}{meV} corresponding to the first interband transition. In addition, we observe a peak emerging at higher energies around \SI{770}{meV}, which aligns with the onset of the second interband transition.

In the field evolution shown in panel (d), one can see weak signatures of the metamagnetic transitions, especially the onset of the saturated state above \SI{3.5}{T}. Nevertheless, the field dependence again follows the monotoous behaviour of the magnetization rather than that of the magnetoresistance.

\subsection{THz magnetoreflectance and magnetotransport}
\begin{figure}[h]
    \centering
    \includegraphics[width=\linewidth]{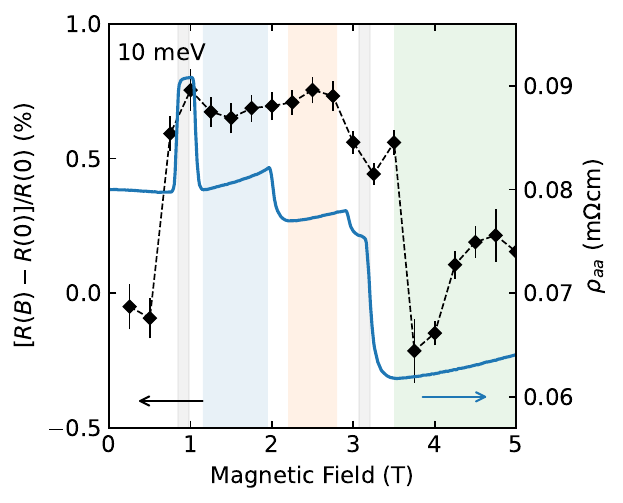}
    \caption{Field dependence of the in-plane magnetoreflectance measured with unpolarized light around 10 meV. To reduce the influence of the spectral noise, the magnetoreflectance spectra were integrated over a flat region between 5 and \SI{12}{meV} to arrive at the ratio of $\Delta R/R$. $\rho_{aa}$ is reproduced from Ref. \cite{Zhao2024}.}
    \label{fig:Tallinn}
\end{figure}
Our results have demonstrated the strong impact of Ho $4f$ electrons on the electronic structure and the optical properties of HoAgGe. Importantly, the Ho $4f$ states form flat bands close to the Fermi level and hybridize with the conduction electrons. Correspondingly, they are expected to also influence the dynamics of the free carriers, i.e. the dc transport and the Drude response. To shed light on this aspect, we analyze the magnetoreflectance spectra at the lowest energies, below \SI{20}{meV}, where the free-carrier response becomes relevant. Due to limitations caused by the sample sizes, we could only extend the magnetoreflectance studies to such low energies for in-plane polarization, displayed in Fig. \ref{fig:Magnetoreflectance}(a). To improve the signal to noise ration, we averaged the signal in a narrow spectral range around \SI{10}{meV}, where it showed no frequency dependence. The resulting field dependence is shown in Fig. \ref{fig:Tallinn}. Upon the first metamagnetic transition, the reflectance exhibits a jump-like increase, followed by only minor variations up to the transition to the saturated state, where a jump-like decrease is observed. For comparison, we plot the field dependence of the in--plane magnetoresistance $\rho_{aa}$ on the second axis and shade the background in the range of the plateau phases. Importantly, the field dependence of the free-carrier response, probed here optically, is completely different from the behaviour of higher-energy charge excitations, probing electronic states further away from the Fermi level.

Let us note that, in this low-frquency region the reflectivity is close to unity, thus the magnetoreflectance is naturally weak. Correspondingly, beyond the error bars representing statistical errors in Fig. \ref{fig:Tallinn}, we cannot exclude systematic errors due to slight field and time dependence in our measurements not coming from the material. Therefore, while the anomalies of the magnetoreflectance clearly originate from the sample, the magnitude of the magnetoreflectance has some uncertainty.

Nevertheless, the obtained field dependence of the THz magnetoreflectance in the Drude regime matches well to that of the dc magnetoresistance, possibly caused by changes in the scattering rate or the effective mass of the free carriers. Indeed, Zhao et al. \cite{Zhao2024} observed negative magnetoresistance in the paramagnetic state just above $T_1$, implying a reduction of scattering from suppressed spin fluctuations in the external field. However, all of the plateau states appear at low temperatures and possess long range magnetic order with little spin fluctuations, hence magnetic scattering on spin fluctuations is expected to be negligible \cite{Zhao2026}. Instead, the sharp features likely reflect band reconstructions close to the Fermi level, modifying the effective mass or the band topology. The latter may mean the emergence or collapse of nodal features near the Fermi level. Along this line, we suspect that the peculiar non-monotonous field-induced changes in the magnetotransport seen by Zhao et al. \cite{Zhao2024} also originate from a modified band structure across the plateau phases.

\section{Discussion}
Our detailed magneto-optical study shows that the polarized optical spectra of HoAgGe can be decomposed into a narrow Drude peak, with a scattering rate in the order of \SI{10}{meV}, and three interband transitions centered around 0.4, 1, and \SI{1.5}{eV}. The first two, also captured by DFT calculations, involve $4f$ states, either directly as unoccupied flat bands serving as excited states or indirectly by splitting itinerant bands from the Ag $4d$, Ag $5p$, and Ge $4p$ states, split by the CEF Coulomb potential projected to the ground $^5I_8$ space of Ho$^{3+}$. This naturally explains the sensitive temperature dependence of $\sigma_{cc}$, since the anisotropy in the $^5I_8$ space diminishes as the temperature increases towards the scale of CEF splittings.

The strong magnetoreflectance observed in the range of the lowest interband transition also supports the scenario above, namely that the interband transitions involve Ho $4f$ states hybridized with other electrons. The Ising anisotropy of the Ho $4f$ moments, leading to the kagome ice rules in this compound, is also the key to understand the influence of the distinct magnetic states on the optical response of HoAgGe. As our most striking observation, the magnetoreflectance in the free-carrier regime follows the unusual non-monotonous field dependence of the dc magnetoresistance, while it mimics the field dependence of the magnetization at the interband excitations. Such a difference between the free-carrier dynamics and interband excitations has been reported for the off-diagonal conductivity, e.g., in the pyrochlore molibdate Nd$_2$Mo$_2$O$_7$ with scalar spin chirality \cite{Kezsmarki2005}, the ferrimagnetic spinel CuCr$_2$Se$_4$ \cite{Bordacs2010}, and SrRuO$_3$ \cite{Fang2003, Demko2012}. This can be understood on a qualitative level as follows. To higher-energy transitions, several bands can contribute thus the influence of topological features in the band structure, being hot spots of Berry curvature, is suppressed. By contrast, interband transitions (Drude response) involve final and excited states from the close vicinity of the Fermi level, thus magnify the contributions from singular points of the band structure, such as (Weyl) Dirac points close to the Fermi energy, generic to the (distorted) kagome lattice \cite{Roychowdhury2024}. In HoAgGe, the Berry phase of low-lying topological states was suggested to be responsible for the observed Hall and magneto-transport response including their peculiar hysteretic behavior \cite{Zhao2024, Bhandari2025}. Distinct variations of the low- and high-energy optical responses across the plateau states, being the key result of this study, support this hypothesis.

Our results elucidate the electronic band structure of HoAgGe and the intimate coupling between the electronic and and spin sectors in this peculiar local moment metal. Since the unoccupied $4f$ flat bands are located in the vicinity of the Fermi energy, their control by different external stimuli, such as carrier doping, pressure or strain, has a huge potential to enhance magnetotransport and magneto-optical responses. Our study reveals a new pathway to control electronic and optical properties by low-lying $4f$ flat bands in rare earth-based itinerant magnets.


\section{Methods}
\subsection{Crystal preparation}
We carried out magneto-reflectance experiments on polished $ab$-cut and $bc$-cut crystals with diameters of 1.4 mm. High-quality single crystals of HoAgGe were grown using the Ag–Ge-rich self-flux method, with typical concentration Ho$_{0.06}$(Ag$_{0.75}$Ge$_{0.25})_{0.94}$ \cite{Morosan2004}. Mixtures were placed in alumina crucibles and sealed in a quartz tube, heated to $1150^\circ$C, held there for 10 h and cooled to $836^\circ$C within 76 h, where the flux was decanted using a centrifuge \cite{Zhao2020}. All single crystals were characterized via X-ray diffraction with a Rigaku X’pert diffractometer using Cu K$\alpha_1$ radiation.

\subsection{FTIR spectroscopy}
The zero-field reflectivity spectra were collected in a Bruker Vertex 80v Fourier-transform spectrometer in the frequency range from 80-25000 cm$^{-1}$ (0.01-3 eV) from room temperature down to 4 K. As references, a silver and gold mirror where used in the mid infrared--visible and far infrared experiments, respectively. The optical conductivity $\sigma(\omega)$ was calculated by Kramers-Kronig transformation of the broadband reflectivity spectra. At this point, the low-energy side of the reflectivity spectrum was extrapolated by using a Hagen-Rubens law and the dc-conductivity, while at the high-energy side the reflectivity spectrum was extrapolated with $\omega^{-1.5}$ for the interband regime and with the free electron behavior setting in at $10^6$\,cm$^{-1}$.

\subsection{Mid- and far-infrared magnetoreflectance}
Relative magneto-reflectance spectra, $[R(B)-R(0)]/R(0)$ in the mid and farinfrared were collected Grenoble in Voigt configuration with the field applied parallel to the $b$ axis in the kagome plane, $H\parallel b$, for orthogonal light polarization ${E}^\omega\parallel a$ and ${E}^\omega\parallel c$. The light of a globar was analyzed by a Bruker Vertex 80v Fourier-transform spectrometer in the energy range between 10 and 600 meV and guided to the sample position by a light-pipe. The sample was cooled to 4\,K and placed in a superconducting coil (SC) for measurements up to 16 T. The reflected light was collected and guided onto an internal bolometer after passing a silicon beam splitter and a wire grid polarizer. The field dependence of the bolometer was corrected with a reference measurement on a gold mirror, evaporated on a silicon substrate.

\subsection{Near-infrared magnetoreflectance}
For the near infrared range, the light of a halogene lamp in a Bruker Vertex80v was coupled to a Hyperion 2000 IR microscope. A mirror objective focused the polarized light into the bore of a 5\,T superconducting Oxford MicrostatMO on an $ac$-cut crystal in Faraday configuration. The reflected light was analyzed by a nitrogen cooled MCT detector. To correct for spectral response not originating from the sample, half of the surface of the crystal was evaporated with a silver layer on which reference measurements for all fields were taken.

\subsection{THz magnetoreflectance}
The THz reflectivity spectra were measured with a Martin-Puplett interferometer and a \SI{0.3}{K} silicon bolometer in Tallinn. The magnetic field was applied in the $ab$ plane of the sample along the $H\parallel b$ direction, while the reflectivity was measured with unpolarized light near normal incidence.

\subsection{Computational Details}
The density functional theory (DFT) calculations were performed using Quantum ESPRESSO \cite{Giannozzi2009, Giannozzi2017} within the GGA+U approximation for the exchange-correlation functional. We have used the projected augmented wave (PAW) pseudopotential of Ho given in Ref.\,\cite{Topsakal2014}, and $U=5.9$ eV, $J=0.7$ eV for Ho according to Ref.\,\cite{Baidak2022}. A $6\times 6\times 8$ $k$-mesh is used for self-consistent calculations. The optical conductivities are calculated using Wannier90 \cite{Mostofi2014} by first interpolating the DFT Hamiltonian from Quantum ESPRESSO on a $3\times 3 \times 4$ $k$-mesh using 72 maximally localized Wannier functions initialized as $s,p$ orbitals on all atoms in the unit cell. We used a lower bound of the disentanglement window of $7.7$ eV (the Fermi energy is $14.07$ eV from DFT), and a frozen window of $[10, 15.1]$ eV. The spread of the Wannier functions is converged to $\sim 2.57$ \AA$^2$ per orbital. The joint DOS (JDOS) and the real part of the longitudinal optical conductivities were calculated using the \texttt{berry} module of Wannier90 and a $k$-mesh of $200\times 200\times 200$.

\section{Data availability}
The IR and magnetoreflectance data are available on Zenodo.

\bibliography{References}

@article{Bhandari2025,
author = {Bhandari, Hari and Chang, Po-Hao and Regmi, Resham Babu and Márkus, Bence G. and Forró, László and Mitchell, John F. and Mazin, Igor I. and Ghimire, Nirmal J.},
title = {{Tunable topological transitions in the frustrated magnet HoAgGe}},
journal = {Communications Materials},
volume = {6},
pages = {52},
year = {2025},
doi = {10.1038/s43246-025-00772-9},
URL = {https://doi.org/10.1038/s43246-025-00772-9},
}

@article{Roychowdhury2024,
author = {Subhajit Roychowdhury  and Kartik Samanta  and Sukriti Singh  and Walter Schnelle  and Yang Zhang  and Jonathan Noky  and Maia G. Vergniory  and Chandra Shekhar  and Claudia Felser },
title = {Enhancement of the anomalous Hall effect by distorting the Kagome lattice in an antiferromagnetic material},
journal = {Proceedings of the National Academy of Sciences},
volume = {121},
number = {30},
pages = {e2401970121},
year = {2024},
doi = {10.1073/pnas.2401970121},
URL = {https://www.pnas.org/doi/abs/10.1073/pnas.2401970121},
}

@Article{Yang2016,
  author    = {Yang, Yongxiang and Walton, Allan and Sheridan, Richard and Güth, Konrad and Gauß, Roland and Gutfleisch, Oliver and Buchert, Matthias and Steenari, Britt-Marie and Van Gerven, Tom and Jones, Peter Tom and Binnemans, Koen},
  journal   = {Journal of Sustainable Metallurgy},
  title     = {{REE Recovery from End-of-Life NdFeB Permanent Magnet Scrap: A Critical Review}},
  year      = {2016},
  issn      = {2199-3831},
  month     = sep,
  number    = {1},
  pages     = {122--149},
  volume    = {3},
  doi       = {10.1007/s40831-016-0090-4},
  publisher = {Springer Science and Business Media LLC},
}

@Article{Simeth_2024,
  author    = {Simeth, W and Rahn, M C and Bauer, A and Meven, M and Pfleiderer, C},
  journal   = {Journal of Physics: Condensed Matter},
  title     = {{Topological aspects of multi-k antiferromagnetism in cubic rare-earth compounds}},
  year      = {2024},
  issn      = {1361-648X},
  month     = feb,
  number    = {21},
  pages     = {215602},
  volume    = {36},
  doi       = {10.1088/1361-648x/ad24bb},
  publisher = {IOP Publishing},
}

@Article{Yin2020,
  author    = {Yin, Jia-Xin and Ma, Wenlong and Cochran, Tyler A. and Xu, Xitong and Zhang, Songtian S. and Tien, Hung-Ju and Shumiya, Nana and Cheng, Guangming and Jiang, Kun and Lian, Biao and Song, Zhida and Chang, Guoqing and Belopolski, Ilya and Multer, Daniel and Litskevich, Maksim and Cheng, Zi-Jia and Yang, Xian P. and Swidler, Bianca and Zhou, Huibin and Lin, Hsin and Neupert, Titus and Wang, Ziqiang and Yao, Nan and Chang, Tay-Rong and Jia, Shuang and Zahid Hasan, M.},
  journal   = {Nature},
  title     = {{Quantum-limit Chern topological magnetism in TbMn$_6$Sn$_6$}},
  year      = {2020},
  issn      = {1476-4687},
  month     = jul,
  number    = {7817},
  pages     = {533--536},
  volume    = {583},
  doi       = {10.1038/s41586-020-2482-7},
  publisher = {Springer Science and Business Media LLC},
}

@Article{Kang2019,
  author    = {Kang, Mingu and Ye, Linda and Fang, Shiang and You, Jhih-Shih and Levitan, Abe and Han, Minyong and Facio, Jorge I. and Jozwiak, Chris and Bostwick, Aaron and Rotenberg, Eli and Chan, Mun K. and McDonald, Ross D. and Graf, David and Kaznatcheev, Konstantine and Vescovo, Elio and Bell, David C. and Kaxiras, Efthimios and van den Brink, Jeroen and Richter, Manuel and Prasad Ghimire, Madhav and Checkelsky, Joseph G. and Comin, Riccardo},
  journal   = {Nature Materials},
  title     = {{Dirac fermions and flat bands in the ideal kagome metal FeSn}},
  year      = {2019},
  issn      = {1476-4660},
  month     = dec,
  number    = {2},
  pages     = {163--169},
  volume    = {19},
  doi       = {10.1038/s41563-019-0531-0},
  publisher = {Springer Science and Business Media LLC},
}

@Article{Ortiz2024,
  author    = {Ortiz, Brenden R. and Sarte, Paul M. and Pokharel, Ganesh and Knudtson, Miles J. and Gomez Alvarado, Steven J. and May, Andrew F. and Calder, Stuart and Mangin-Thro, Lucile and Wildes, Andrew R. and Zhou, Haidong and Sala, Gabriele and Wiebe, Chris R. and Wilson, Stephen D. and Paddison, Joseph A. M. and Aczel, Adam A.},
  journal   = {Physical Review B},
  title     = {{Revisiting spin ice physics in the ferromagnetic Ising pyrochlore Pr$_2$Sn$_2$O$_7$}},
  year      = {2024},
  issn      = {2469-9969},
  month     = apr,
  number    = {13},
  pages     = {134420},
  volume    = {109},
  doi       = {10.1103/physrevb.109.134420},
  publisher = {American Physical Society (APS)},
}

@Article{Singh_2008,
  author    = {Singh, Surjeet and Suryanarayanan, R. and Tackett, R. and Lawes, G. and Sood, A. K. and Berthet, P. and Revcolevschi, A.},
  journal   = {Physical Review B},
  title     = {{Ordered spin-ice state in the geometrically frustrated metallic ferromagnet Sm$_2$Mo$_4$O$_7$}},
  year      = {2008},
  issn      = {1550-235X},
  month     = jan,
  number    = {2},
  pages     = {020406},
  volume    = {77},
  doi       = {10.1103/physrevb.77.020406},
 publisher = {American Physical Society (APS)},
}

@Article{Yasui2001,
  author    = {Yasui, Yukio and Kondo, Yasuyuki and Kanada, Masaki and Ito, Masafumi and Harashina, Hiroshi and Sato, Masatoshi and Kakurai, Kazuhisa},
  journal   = {Journal of the Physical Society of Japan},
  title     = {{Magnetic Structure of Nd$_2$Mo$_2$O$_7$}},
  year      = {2001},
  issn      = {1347-4073},
  month     = jan,
  number    = {1},
  pages     = {284--289},
  volume    = {70},
  doi       = {10.1143/jpsj.70.284},
  publisher = {Physical Society of Japan},
}

@Article{Hirschberger2019,
  author    = {Hirschberger, Max and Nakajima, Taro and Gao, Shang and Peng, Licong and Kikkawa, Akiko and Kurumaji, Takashi and Kriener, Markus and Yamasaki, Yuichi and Sagayama, Hajime and Nakao, Hironori and Ohishi, Kazuki and Kakurai, Kazuhisa and Taguchi, Yasujiro and Yu, Xiuzhen and Arima, Taka-hisa and Tokura, Yoshinori},
  journal   = {Nature Communications},
  title     = {{Skyrmion phase and competing magnetic orders on a breathing kagomé lattice}},
  year      = {2019},
  issn      = {2041-1723},
  pages     = {5831},
  month     = dec,
  number    = {1},
  volume    = {10},
  doi       = {10.1038/s41467-019-13675-4},
  publisher = {Springer Science and Business Media LLC},
}

@article{Dieke1963,
author = {G. H. Dieke and H. M. Crosswhite},
journal = {Appl. Opt.},
number = {7},
pages = {675--686},
publisher = {Optica Publishing Group},
title = {{The Spectra of the Doubly and Triply Ionized Rare Earths}},
volume = {2},
month = {Jul},
year = {1963},
url = {https://opg.optica.org/ao/abstract.cfm?URI=ao-2-7-675},
doi = {10.1364/AO.2.000675},
}

@article{Kezsmarki2005,
  title = {{Magneto-optical effect induced by spin chirality of the itinerant ferromagnet ${\mathrm{Nd}}_{2}{\mathrm{Mo}}_{2}{\mathrm{O}}_{7}$}},
  author = {K\'ezsm\'arki, I. and Onoda, S. and Taguchi, Y. and Ogasawara, T. and Matsubara, M. and Iguchi, S. and Hanasaki, N. and Nagaosa, N. and Tokura, Y.},
  journal = {Phys. Rev. B},
  volume = {72},
  issue = {9},
  pages = {094427},
  numpages = {6},
  year = {2005},
  month = {Sep},
  publisher = {American Physical Society},
  doi = {10.1103/PhysRevB.72.094427},
  url = {https://link.aps.org/doi/10.1103/PhysRevB.72.094427}
}

@article{Morosan2004,
    author = {Morosan, E. and Bud'ko, S. L. and Canfield, P. C. and Torikachvili, M. S. and Lacerda, A. H.},
    doi = {10.1016/j.jmmm.2003.11.014},
    issn = {03048853},
    journal = {Journal of Magnetism and Magnetic Materials},
    number = {3},
    pages = {298--321},
    title = {{Thermodynamic and transport properties of RAgGe (R=Tb-Lu) single crystals}},
    volume = {277},
    year = {2004}
}

@article{Zhao2020,
    author = {Zhao, Kan and Deng, Hao and Chen, Hua and Ross, Kate A and Petř{\'{i}}{\v{c}}ek, Vaclav and G{\"{u}}nther, Gerrit and Russina, Margarita and Hutanu, Vladimir and Gegenwart, Philipp},
    doi = {10.1126/science.aaw1666},
    issn = {0036-8075},
    journal = {Science},
    month = {3},
    number = {6483},
    pages = {1218--1223},
    title = {{Realization of the kagome spin ice state in a frustrated intermetallic compound}},
    url = {https://www.science.org/doi/10.1126/science.aaw1666},
    volume = {367},
    year = {2020}
}

@article{Zhao2026,
  title = {Three-Dimensional XY Universality and Nonlinear Magnetic Susceptibility in a Kagome Ice Compound},
  author = {Zhao, Kan and Deng, Hao and Chen, Hua and Ma, Nvsen and Oefele, Noah and Guo, Jiesen and Cui, Xueling and Tang, Chen and Gutmann, Matthias J. and Mueller, Thomas and Su, Yixi and Hutanu, Vladimir and Jin, Changqing and Gegenwart, Philipp},
  journal = {Phys. Rev. X},
  volume = {16},
  issue = {2},
  pages = {021043},
  numpages = {26},
  year = {2026},
  month = {May},
  publisher = {American Physical Society},
  doi = {10.1103/xl5f-zj9p},
  url = {https://link.aps.org/doi/10.1103/xl5f-zj9p}
}

@article{Wills2002,
    author = {Wills, A. S. and Ballou, R. and Lacroix, C.},
    doi = {10.1103/PhysRevB.66.144407},
    issn = {1550235X},
    journal = {Physical Review B},
    number = {14},
    pages = {1--6},
    title = {{Model of localized highly frustrated ferromagnetism: The kagom{\'{e}} spin ice}},
    volume = {66},
    year = {2002}
}

@article{Li2022,
    author = {Li, N. and Huang, Q. and Yue, X. Y. and Guang, S. K. and Xia, K. and Wang, Y. Y. and Li, Q. J. and Zhao, X. and Zhou, H. D. and Sun, X. F.},
    doi = {10.1103/PhysRevB.106.014416},
    issn = {24699969},
    journal = {Physical Review B},
    number = {1},
    pages = {1--8},
    title = {{Low-temperature transport properties of the intermetallic compound HoAgGe with a kagome spin-ice state}},
    volume = {106},
    year = {2022}
}

@Article{Zhao2024,
  	author    = {Zhao, K. and Tokiwa, Y. and Chen, H. and Gegenwart, P.},
  	journal   = {Nature Physics},
  	title     = {{Discrete degeneracies distinguished by the anomalous Hall effect in a metallic kagome ice compound}},
  	year      = {2024},
  	issn      = {1745-2481},
  	month     = {1},
  	number    = {3},
  	pages     = {442--449},
  	volume    = {20},
  	doi       = {10.1038/s41567-023-02307-w},
  	publisher = {Springer Science and Business Media LLC},
}

@Article{Du2022,
	author    = {Du, Qianheng and Hu, Zhixiang and Han, Myung-Geun and Camino, Fernando and Zhu, Yimei and Petrovic, C.},
	journal   = {Phys. Rev. Lett.},
	title     = {{Topological Hall Effect Anisotropy in Kagome Bilayer Metal}},
	year      = {2022},
	issn      = {1079-7114},
	month     = nov,
	number    = {23},
	pages     = {236601},
	volume    = {129},
	doi       = {10.1103/physrevlett.129.236601},
	publisher = {American Physical Society (APS)},
}

@Article{Yang2022,
	author    = {Yang, T. Y. and Wan, Q. and Song, J. P. and Du, Z. and Tang, J. and Wang, Z. W. and Plumb, N. C. and Radovic, M. and Wang, G. W. and Wang, G. Y. and Sun, Z. and Yin, Jia-Xin and Chen, Z. H. and Huang, Y. B. and Yu, R. and Shi, M. and Xiong, Y. M. and Xu, N.},
	journal   = {Quantum Frontiers},
	title     = {{Fermi-level flat band in a kagome magnet}},
	year      = {2022},
	issn      = {2731-6106},
	month     = nov,
	number    = {1},
	pages     = {14},
	volume    = {1},
	doi       = {10.1007/s44214-022-00017-7},
	publisher = {Springer Science and Business Media LLC},
}

@Article{Sales2019,
	author    = {Sales, Brian C. and Yan, Jiaqiang and Meier, William R. and Christianson, Andrew D. and Okamoto, Satoshi and McGuire, Michael A.},
	journal   = {Physical Review Materials},
	title     = {{Electronic, magnetic, and thermodynamic properties of the kagome layer compound FeSn}},
	year      = {2019},
	issn      = {2475-9953},
	month     = nov,
	number    = {11},
	pages     = {114203},
	volume    = {3},
	doi       = {10.1103/physrevmaterials.3.114203},
	publisher = {American Physical Society (APS)},
}

@Article{EbadAllah2024,
	author    = {Ebad-Allah, J. and Jiang, M.-C. and Borkenhagen, R. and Meggle, F. and Prodan, L. and Tsurkan, V. and Schilberth, F. and Guo, G.-Y. and Arita, R. and Kézsmárki, I. and Kuntscher, C. A.},
	journal   = {Physical Review B},
	title     = {{Optical anisotropy of the kagome magnet FeSn: Dominant role of excitations between kagome and Sn layers}},
	year      = {2024},
	issn      = {2469-9969},
	month     = {5},
	number    = {20},
	pages     = {l201106},
	volume    = {109},
	doi       = {10.1103/physrevb.109.l201106},
	publisher = {American Physical Society (APS)},
}

@Article{Topsakal2014,
  author           = {Topsakal, M. and Wentzcovitch, R.M.},
  journal          = {Computational Materials Science},
  title            = {{Accurate projected augmented wave (PAW) datasets for rare-earth elements (RE=La–Lu)}},
  year             = {2014},
  issn             = {0927-0256},
  month            = dec,
  pages            = {263--270},
  volume           = {95},
  creationdate     = {2025-04-11T16:36:21},
  doi              = {10.1016/j.commatsci.2014.07.030},
  modificationdate = {2025-04-11T16:49:59},
  publisher        = {Elsevier BV},
}

@Article{Baidak2022,
  author           = {Baidak, Semyon T. and Lukoyanov, Alexey V.},
  journal          = {Materials},
  title            = {{Common Topological Features in Band Structure of RNiSb and RSb Compounds for R = Tb, Dy, Ho}},
  year             = {2022},
  issn             = {1996-1944},
  month            = dec,
  number           = {1},
  pages            = {242},
  volume           = {16},
  creationdate     = {2025-04-11T16:36:47},
  doi              = {10.3390/ma16010242},
  modificationdate = {2025-04-11T16:49:35},
  publisher        = {MDPI AG},
}

@Article{Giannozzi2017,
  author           = {Giannozzi, P and Andreussi, O and Brumme, T and Bunau, O and Buongiorno Nardelli, M and Calandra, M and Car, R and Cavazzoni, C and Ceresoli, D and Cococcioni, M and Colonna, N and Carnimeo, I and Dal Corso, A and de Gironcoli, S and Delugas, P and DiStasio, R A and Ferretti, A and Floris, A and Fratesi, G and Fugallo, G and Gebauer, R and Gerstmann, U and Giustino, F and Gorni, T and Jia, J and Kawamura, M and Ko, H-Y and Kokalj, A and Küçükbenli, E and Lazzeri, M and Marsili, M and Marzari, N and Mauri, F and Nguyen, N L and Nguyen, H-V and Otero-de-la-Roza, A and Paulatto, L and Poncé, S and Rocca, D and Sabatini, R and Santra, B and Schlipf, M and Seitsonen, A P and Smogunov, A and Timrov, I and Thonhauser, T and Umari, P and Vast, N and Wu, X and Baroni, S},
  journal          = {Journal of Physics: Condensed Matter},
  title            = {{Advanced capabilities for materials modelling with Quantum ESPRESSO}},
  year             = {2017},
  issn             = {1361-648X},
  month            = oct,
  number           = {46},
  pages            = {465901},
  volume           = {29},
  creationdate     = {2025-04-11T16:37:29},
  doi              = {10.1088/1361-648x/aa8f79},
  modificationdate = {2025-04-11T16:49:45},
  publisher        = {IOP Publishing},
}

@Article{Giannozzi2009,
  author           = {Giannozzi, Paolo and Baroni, Stefano and Bonini, Nicola and Calandra, Matteo and Car, Roberto and Cavazzoni, Carlo and Ceresoli, Davide and Chiarotti, Guido L and Cococcioni, Matteo and Dabo, Ismaila and Dal Corso, Andrea and de Gironcoli, Stefano and Fabris, Stefano and Fratesi, Guido and Gebauer, Ralph and Gerstmann, Uwe and Gougoussis, Christos and Kokalj, Anton and Lazzeri, Michele and Martin-Samos, Layla and Marzari, Nicola and Mauri, Francesco and Mazzarello, Riccardo and Paolini, Stefano and Pasquarello, Alfredo and Paulatto, Lorenzo and Sbraccia, Carlo and Scandolo, Sandro and Sclauzero, Gabriele and Seitsonen, Ari P and Smogunov, Alexander and Umari, Paolo and Wentzcovitch, Renata M},
  journal          = {Journal of Physics: Condensed Matter},
  title            = {{QUANTUM ESPRESSO: a modular and open-source software project for quantum simulations of materials}},
  year             = {2009},
  issn             = {1361-648X},
  month            = sep,
  number           = {39},
  pages            = {395502},
  volume           = {21},
  creationdate     = {2025-04-11T16:37:46},
  doi              = {10.1088/0953-8984/21/39/395502},
  modificationdate = {2025-04-11T16:50:06},
  publisher        = {IOP Publishing},
}

@Article{Mostofi2014,
  author           = {Mostofi, Arash A. and Yates, Jonathan R. and Pizzi, Giovanni and Lee, Young-Su and Souza, Ivo and Vanderbilt, David and Marzari, Nicola},
  journal          = {Computer Physics Communications},
  title            = {{An updated version of wannier90: A tool for obtaining maximally-localised Wannier functions}},
  year             = {2014},
  issn             = {0010-4655},
  month            = aug,
  number           = {8},
  pages            = {2309--2310},
  volume           = {185},
  creationdate     = {2025-04-11T16:38:21},
  doi              = {10.1016/j.cpc.2014.05.003},
  modificationdate = {2025-04-11T16:49:53},
  publisher        = {Elsevier BV},
}

@article{Bordacs2010,
  title={{Experimental band structure of the nearly half-metallic CuCr$_2$Se$_4$: an optical and magneto-optical study}},
  author={Bord{\'a}cs, S{\'a}ndor and K{\'e}zsm{\'a}rki, Istv{\'a}n and Ohgushi, K and Tokura, Y},
  journal={New Journal of Physics},
  volume={12},
  number={5},
  pages={053039},
  year={2010},
  publisher={IOP Publishing},
  doi = {10.1088/1367-2630/12/5/053039},
  url = {https://iopscience.iop.org/article/10.1088/1367-2630/12/5/053039}
}

@article{Fang2003,
author = {Zhong Fang  and Naoto Nagaosa  and Kei S. Takahashi  and Atsushi Asamitsu  and Roland Mathieu  and Takeshi Ogasawara  and Hiroyuki Yamada  and Masashi Kawasaki  and Yoshinori Tokura  and Kiyoyuki Terakura },
title = {The Anomalous Hall Effect and Magnetic Monopoles in Momentum Space},
journal = {Science},
volume = {302},
number = {5642},
pages = {92-95},
year = {2003},
doi = {10.1126/science.1089408},
URL = {https://www.science.org/doi/abs/10.1126/science.1089408}}

@article{Demko2012,
  title = {{Disorder Promotes Ferromagnetism: Rounding of the Quantum Phase Transition in ${\mathrm{Sr}}_{1\ensuremath{-}x}{\mathrm{Ca}}_{x}{\mathrm{RuO}}_{3}$}},
  author = {Demk\'o, L. and Bord\'acs, S. and Vojta, T. and Nozadze, D. and Hrahsheh, F. and Svoboda, C. and D\'ora, B. and Yamada, H. and Kawasaki, M. and Tokura, Y. and K\'ezsm\'arki, I.},
  journal = {Phys. Rev. Lett.},
  volume = {108},
  issue = {18},
  pages = {185701},
  numpages = {5},
  year = {2012},
  month = {May},
  publisher = {American Physical Society},
  doi = {10.1103/PhysRevLett.108.185701},
  url = {https://link.aps.org/doi/10.1103/PhysRevLett.108.185701}
}

@article{Kurumaji2019,
author = {Takashi Kurumaji  and Taro Nakajima  and Max Hirschberger  and Akiko Kikkawa  and Yuichi Yamasaki  and Hajime Sagayama  and Hironori Nakao  and Yasujiro Taguchi  and Taka-hisa Arima  and Yoshinori Tokura },
title = {Skyrmion lattice with a giant topological Hall effect in a frustrated triangular-lattice magnet},
journal = {Science},
volume = {365},
number = {6456},
pages = {914-918},
year = {2019},
doi = {10.1126/science.aau0968},
URL = {https://www.science.org/doi/abs/10.1126/science.aau0968}
}

@article{Taguchi2001,
author = {Y. Taguchi  and Y. Oohara  and H. Yoshizawa  and N. Nagaosa  and Y. Tokura },
title = {Spin Chirality, Berry Phase, and Anomalous Hall Effect in a Frustrated Ferromagnet},
journal = {Science},
volume = {291},
number = {5513},
pages = {2573-2576},
year = {2001},
doi = {10.1126/science.1058161},
URL = {https://www.science.org/doi/abs/10.1126/science.1058161}
}

@Article{Khanh2020,
author={Khanh, Nguyen Duy
and Nakajima, Taro
and Yu, Xiuzhen
and Gao, Shang
and Shibata, Kiyou
and Hirschberger, Max
and Yamasaki, Yuichi
and Sagayama, Hajime
and Nakao, Hironori
and Peng, Licong
and Nakajima, Kiyomi
and Takagi, Rina
and Arima, Taka-hisa
and Tokura, Yoshinori
and Seki, Shinichiro},
title={Nanometric square skyrmion lattice in a centrosymmetric tetragonal magnet},
journal={Nature Nanotechnology},
year={2020},
month={Jun},
day={01},
volume={15},
number={6},
pages={444-449},
issn={1748-3395},
doi={10.1038/s41565-020-0684-7},
url={https://doi.org/10.1038/s41565-020-0684-7}
}

@article{Bramwell2001,
author = {Steven T. Bramwell  and Michel J. P. Gingras },
title = {Spin Ice State in Frustrated Magnetic Pyrochlore Materials},
journal = {Science},
volume = {294},
number = {5546},
pages = {1495-1501},
year = {2001},
doi = {10.1126/science.1064761},
URL = {https://www.science.org/doi/abs/10.1126/science.1064761}
}

@article{Morris2009,
author = {D. J. P. Morris  and D. A. Tennant  and S. A. Grigera  and B. Klemke  and C. Castelnovo  and R. Moessner  and C. Czternasty  and M. Meissner  and K. C. Rule  and J.-U. Hoffmann  and K. Kiefer  and S. Gerischer  and D. Slobinsky  and R. S. Perry },
title = {{Dirac Strings and Magnetic Monopoles in the Spin Ice Dy$_2$Ti$_2$O$_7$}},
journal = {Science},
volume = {326},
number = {5951},
pages = {411-414},
year = {2009},
doi = {10.1126/science.1178868},
URL = {https://www.science.org/doi/abs/10.1126/science.1178868},
}

@article{Chen2025,
  title = {Topological Hall Effect of Skyrmions from first Principles},
  author = {Chen, Hsiao-Yi and Nomoto, Takuya and Hirschberger, Max and Arita, Ryotaro},
  journal = {Phys. Rev. X},
  volume = {15},
  issue = {1},
  pages = {011054},
  numpages = {16},
  year = {2025},
  month = {Mar},
  publisher = {American Physical Society},
  doi = {10.1103/PhysRevX.15.011054},
  url = {https://link.aps.org/doi/10.1103/PhysRevX.15.011054}
}

\section{Funding}
This work was supported by the Hungarian National Research, Development and Innovation Office – NKFIH, ADVANCED 152684 and by the bilateral program of the Estonian and Hungarian Academies of Sciences under the contract No.~NKM2025-25/2025 and by the Deutsche Forschungsgemeinschaft (DFG, German Research Foundation) – TRR 360 – 492547816. The work in Beijing was supported by the National Key R\&D Program of China (Grant No. 2023YFA1406003), the Beijing Natural Science Foundation (Grant No. JQ24012), National Natural Science Foundation of China (Grants No. 12274015), and the Fundamental Research Funds for the Central Universities. LD and HC acknowledge support by U.S. National Science Foundation grants \#1945023 and \#2531960. The authors also acknowledge the support of the LNCMI-CNRS in Grenoble, a member of the European Magnetic Field Laboratory (EMFL).

\section{Author Contributions}
FS performed the reflectivity measurements and the data analysis. FS, NH, UN and TR collected the THz magnetoreflectance data. FS, SB, FL, IM and MO. measured the FIR/MIR magnetoreflectance data. FS measured the NIR magnetoreflectance spectra. JG and KZ grew and prepared the samples. LD and HC performed the \textit{ab initio} calculations. FS, KZ, PG, HC, IK and SB contributed to data interpretation and discussions. FS, HC, IK and SB wrote the paper. IK and SB planned and coordinated the project. 

\section{Competing Interests}
The authors declare no competing interests.

\clearpage

\end{document}